**Two-phase coexistence in the hard-disk model**

**Leslie V Woodcock**

Department of Chemical and Biomolecular Engineering
National University of Singapore, Singapore 117576
e-mail chewlv@nus.edu.sg

**A two-dimensional system of 10000 hard disks with square periodic boundary conditions, at a density in the middle of the 2-phase region predicted from equation-of-state data, when subjected to a weak external uniform force, is seen to phase separate. Thermodynamic profiles of the inhomogeneous two-phase system agree with the local density approximation (LDA) of density functional theory. There is no indication of any mesophase.**

It is well known that there is only one phase in a system of one-dimensional hard lines, from the low density ideal gas limit to the maximum packing. “Phase” implies that thermodynamic properties such as molar volume and entropy are single valued state functions of temperature at constant pressure, and continuous in all their derivatives. For three dimensional hard spheres, there are two equilibrium phases; a fluid phase, a crystalline phase and a first-order thermodynamic phase transition at the freezing point. The number of phases in two dimensions, however, has been a matter of some controversy. Some investigators have suggested two phases, whereas others infer no first–order discontinuity in entropy and density, but rather, an intermediate third phase which has been described as the “hexatic mesophase” [see e.g. ref. 1] .

From the known virial coefficients [2], an analytic closed-form equation-of-state for the hard-disk fluid, up to the density of the first phase transition has recently been derived [3]

$$Z=1+\sum_{n=2}^{m}B_n\rho^{*(n-1)}+\rho^{*m}\left[\frac{(C-mA)}{(1-\rho^*)}-\frac{A}{(1-\rho^*)^2}\right] \qquad (1)$$

where $\rho^* = \rho/\rho_0$ ; values for the two constants C and A are C = 5.4995 and A = 0.1125 and m = 11. The virial coefficients are given in references [2] and [3]. This closed virial equation-of-state is extremely accurate up to a pressure about ¾ of the coexistence pressure at the phase transition, whereupon it begins to deviate (Figure 1) [3]. Whether the transition is either first order or not, it seems likely that the thermodynamic system may be exhibiting “pre-transition” effects which are commonly associated with weak first-order or second-order phase transitions, but not reflected in the virial expansion.

Alder et al.[4] have parameterized the equation-of-state of the 2D crystal in an expansion in powers of a "free area" $\alpha$ = $(1/\rho^*-1)$, analogous to a free volume for spheres. The crystal expansion can be written

$$Z = 2/\alpha + \sum_{n=1} C_n \, \alpha^{(n-1)} \qquad (2)$$

where $Z = p/\rho k_B T$.

New MD computations for 10000 spheres have been undertaken in the crystal phase and the data obtained has been parameterized up to the point of the phase transition with this expansion truncated at $C_3$ with the values $C_1$= 1.945, $C_2$= 0.500 and $C_3$ = 2.985. Note that these parameters are slightly different from those given by Alder et al. obtained for smaller systems with perfect order in hexagonal boundary conditions. At thermodynamic equilibrium, the crystal phase contains vacancy and dislocation defects that may be suppressed in the small hexagonal systems.

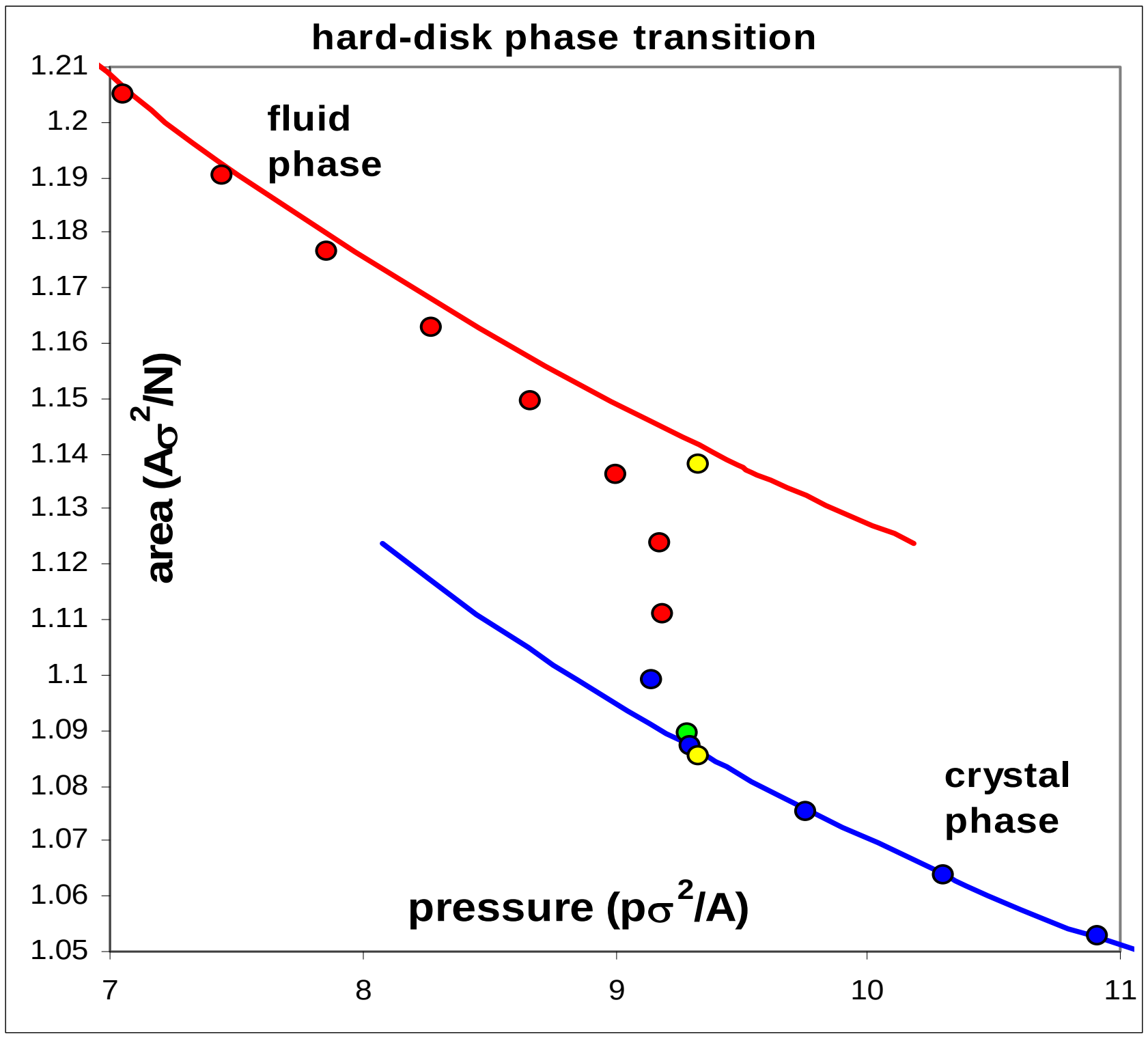

**Figure 1:** Equation-of-state for the hard-disk fluid in the vicinity of the phase transition showing the crystal equation-of-state (blue line), fluid equation-of-state (red line) MD results of Kolafa and Rottner [2] (red circles), and present MD data (N=10000) (blue circles); also plotted (green circle) is the state point from Jaster [1] for N= $(1024)^2$, and the Hoover-Ree [5] fluid freezing and crystal melting points (yellow circles) at their calculated coexistence pressure (9.33).

In order to test whether a point in the two-phase region is homogeneous, as implied by a hexatic mesophase, or heterogeneous as expected form classical thermodynamics for a first-order phase transition, a weak external field has been applied. The objective of this computer experiment is to see whether any phase separation occurs. The state point chosen for this experiment is $\rho\sigma^2$= 0.9 (reduced area $A/N\sigma^2$= 1.1111) i.e. intermediate between the predicted fluid freezing density and crystal melting density from chemical potential calculations [5]. A system of 10000 disks with square boundaries was brought to equilibrium by MD simulation. An equilibrated configuration is shown below (Figure 2). A close inspection, shows heterogeneities, rather difficult to see at first, with nebulous boundaries, on a distance scale of the order 10σ.

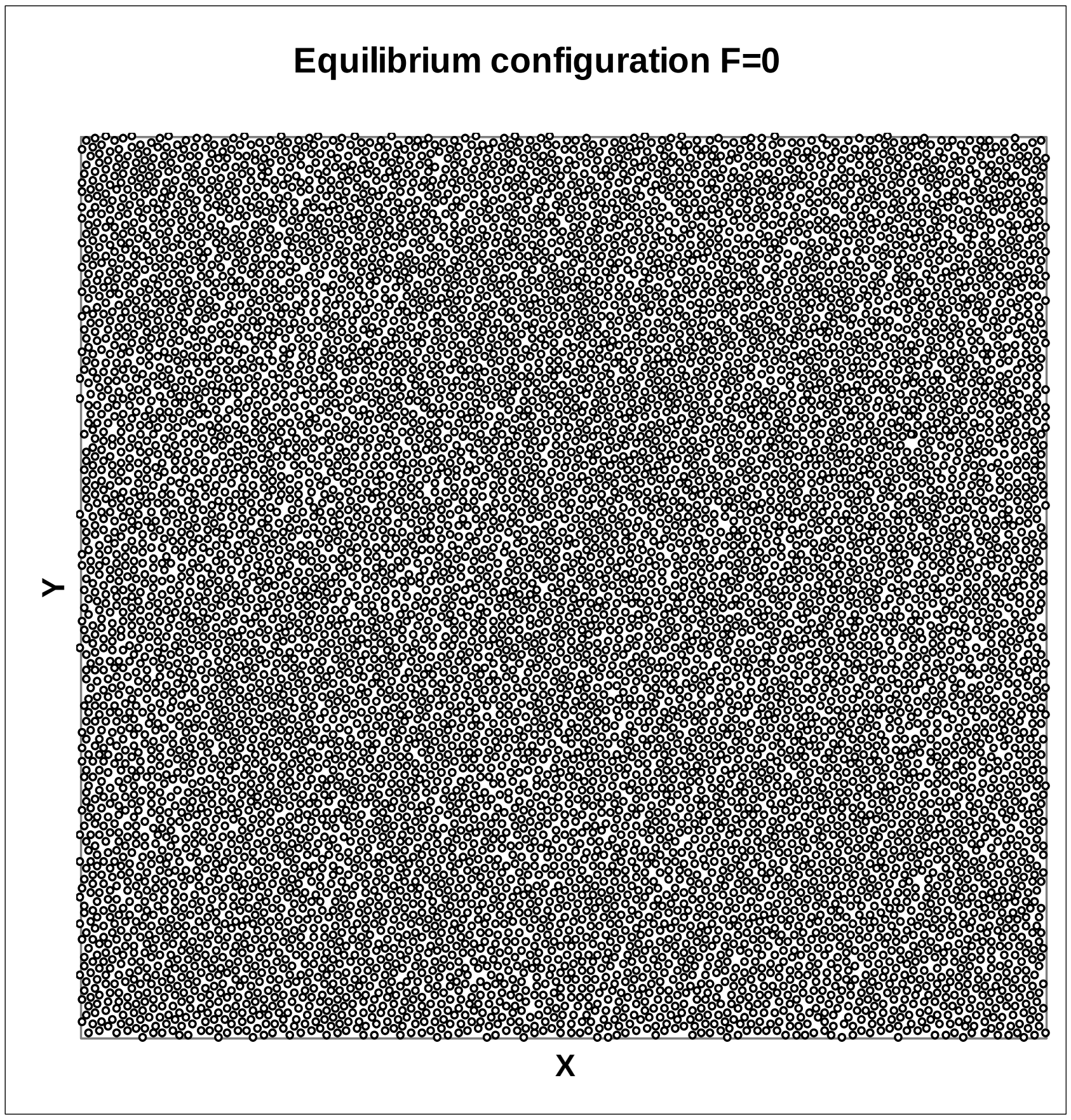

**Figure 2**: A configuration of 10000 hard disks, with square periodic boundary conditions equilibrated at the reduced density $\rho\sigma^2$ = 0.9 ( $A\sigma^2/ N$ = 1.1111) i.e. at a state point intermediate between the fluid density and the crystal density at the coexistence pressure calculated from the point of equal chemical potential from the respective equations-of-state of the fluid and crystal phases.

After equilibration is established a weak external uniform field (F) is applied in the Y-direction. The periodic boundary condition in the Y-component is then replaced by a rigid reflecting wall at Y= 0 to counterbalance the external force. When a steady-state is reached, equilibration is re-established in the inhomogeneous system. All the thermodynamic properties become functions of height (Y), equipartition of energy prevails, and the temperature (T) remains uniform at $k_BT/ <mv^2> = 1.0$

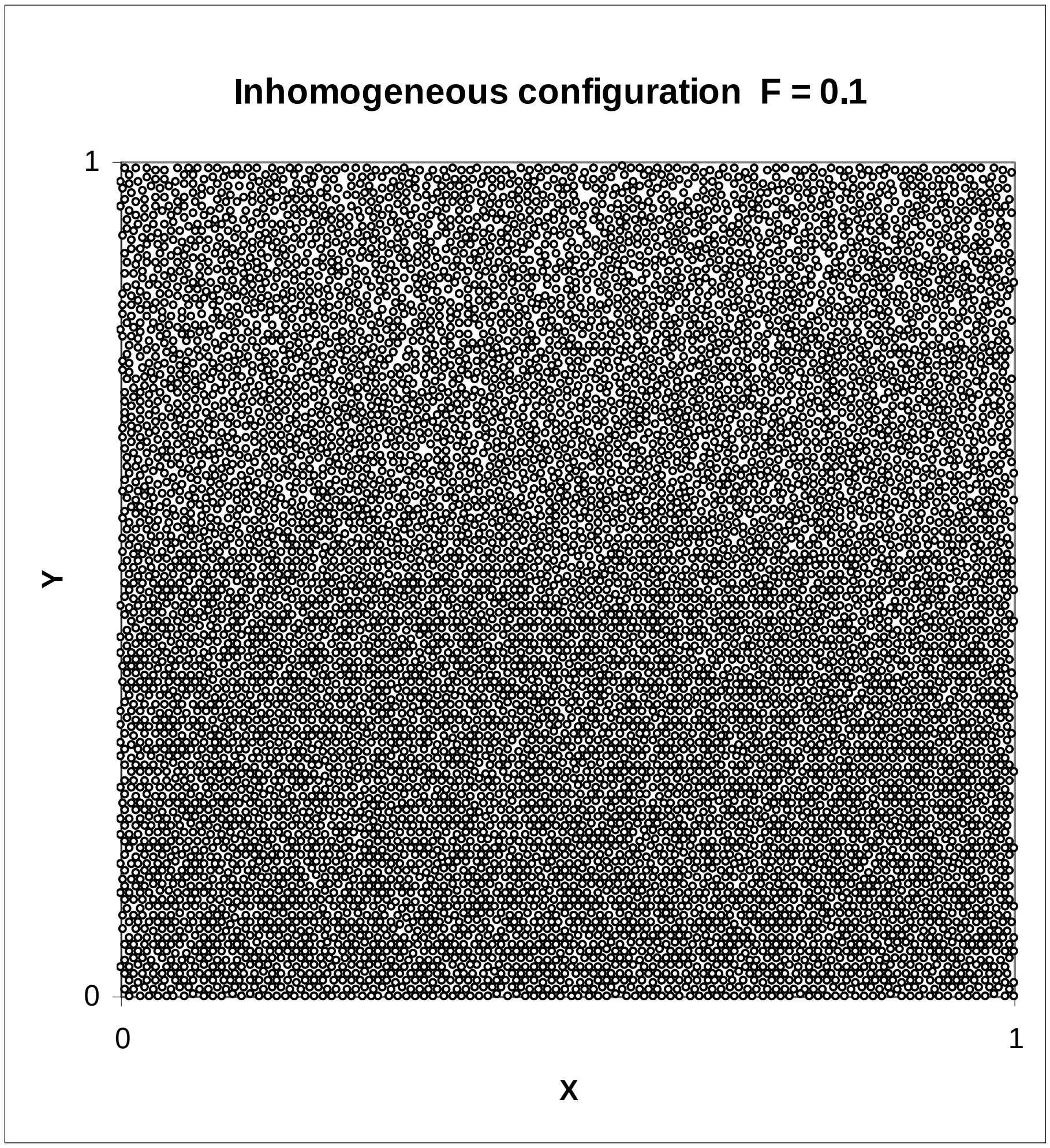

**Figure 3**: The same system as in figure 2 re-equilibrated in the presence of an external field F = 0.1 ($k_BT/ \sigma$), where $k_BT$ is the mean kinetic energy of the system per particle. The configuration shows that the system has phase separated into the two phases, with the higher density crystal phase coexisting below the fluid phase, with an interface between them. There is no evidence of a third intermediate phase.

Profiles of all the thermodynamic properties, density, temperature, and pressures, are obtained. The pressure as calculated from the virial theorem is resolved into its two components $p_{xx}$ and $p_{yy}$, both of which are obtained as a function of Y. The point of contact of the collision that exchanges momentum is the Y-value to which that collision contributes. Profiles of density and pressure, and stress ($p_{xx}$-$p_{yy}$) which determines the interfacial tension, are shown in Figure 4.

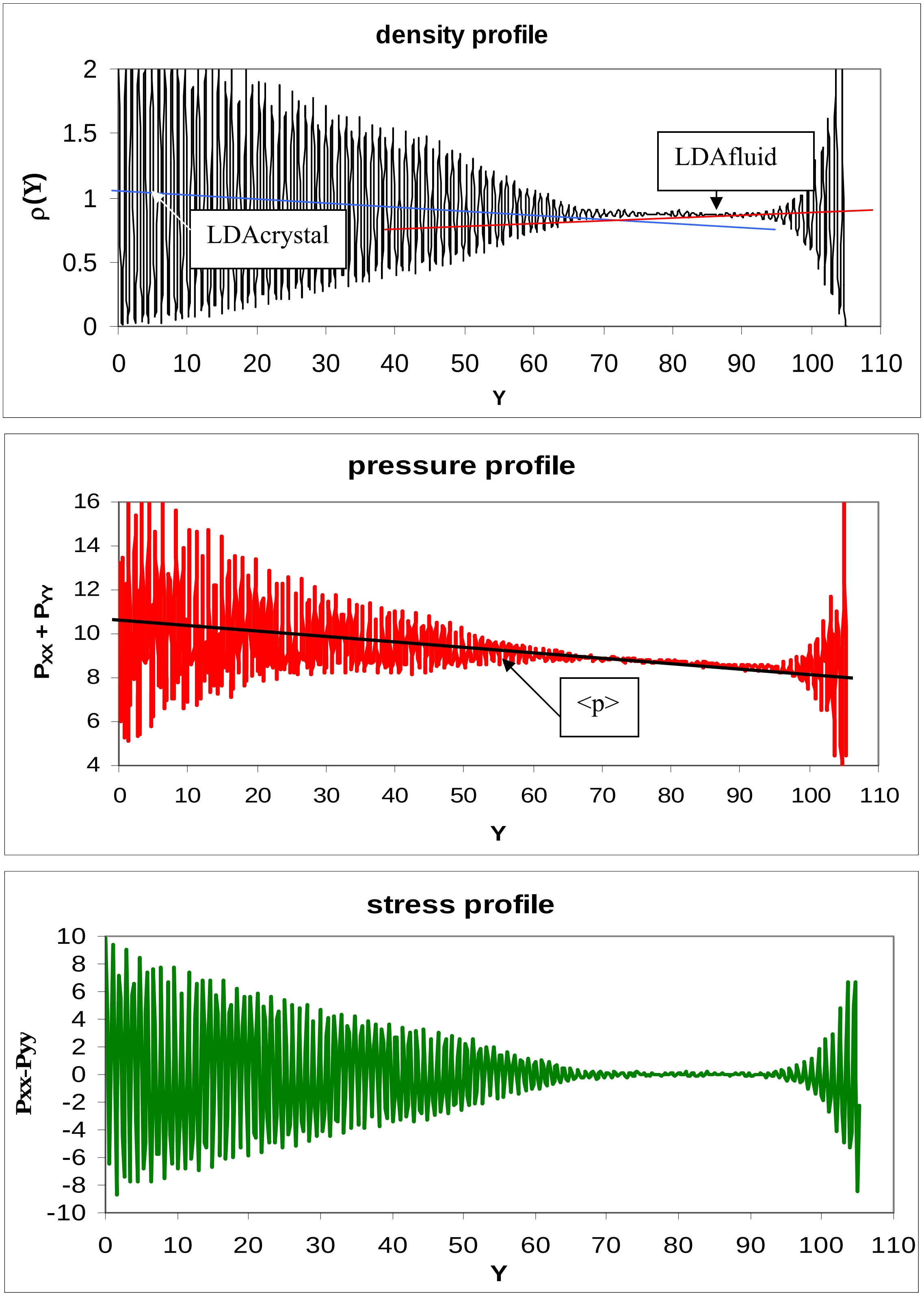

**Figure 4**: Profiles of density, total pressure ($p_{xx} + p_{yy}$) and stress ($p_{xx} - p_{yy}$) for the inhomogeneous system of 10000 hard disks subjected to a weak external uniform force.

All the profiles quantitatively confirm the conclusion seen in Figure 3, i.e. that the inhomogeneous system has 2 phases. Moreover, the thermodynamic properties of the two phases in coexistence in the inhomogeneous system can be shown to be consistent with the thermodynamic description of each phase on either side of the interface.

The data of the thermodynamic profiles in Figure 4 can be used to test the tenets of the local density approximation (LDA) of density functional theory in two dimensions. The LDA approximation has been shown to be accurate for first-order two-phase crystal-fluid coexistence in simple 3-dimensional systems [6]. At any local point in the profile the density of an inhomogeneous fluid is predicted from the equation-of-state at the same temperature and pressure of the equilibrium homogeneous fluid. The blue and red lines are the LDA predictions for the crystal and fluid respectively, on the density profile in Figure 4 as obtained from the smoothed mean pressure profile are in good agreement with the inhomogeneous MD data on their respective sides of the phase transition around the coexistence pressure $p\sigma^2/k_BT = 9.3$.

Also shown in Figure 4 is the profile of the normal pressure difference, i.e. $(p_{yy} - p_{xx})$. This stress is only none-zero in an inhomogeneous system at an interface, and is everywhere zero for a homogeneous phase. There is no sign of a peak at the interface between the two phases which would be indicative of a significant surface tension. This suggests that the interfacial tension in this 2-D system, unlike the 3-D counterpart, is extremely small; so small as to be undetectable during extensive averaging. This observation is consistent with the apparent absence of any kinetic barrier to crystallization by homogeneous nucleation in two-dimensional fluids.

The first-order phase transition, as evidenced by the data in Figures 1 and 4, has associated with it an unusual asymmetry: only the fluid side shows the anomalous pre-transition deviation of the MD data from the equation-of-state. The description of the phase transition, therefore, remains a mystery. If the virial equation does not represent the equilibrium fluid in the intermediate density region leading up to the phase transition, the question arises, what does? Could there possibly be a third phase, one wonders, not the "hexagonal mesophase", but a second fluid phase that has its origins at the fluid-fluid percolation transition [7] ?

---